\documentstyle[preprint,aps]{revtex}
\begin{document}
\tighten
\draft
\title {
Non-equilibrium noise in a mesoscopic conductor:\\
microscopic analysis }
\author{
B. L. Altshuler$^a$, L. S. Levitov$^{a,b}$, A. Yu. Yakovets$^b$}
\address
{ (a) Massachusetts Institute of Technology, 12-112,
77 Massachusetts Ave., Cambridge, MA 02139 \\
(b) L.D.Landau Institute for Theoretical Physics,
2, Kosygin str., Moscow 117334, Russia}

\maketitle

\begin{abstract}
Current fluctuations are studied in a mesoscopic conductor using
non-equilibrium Keldysh technique. We derive a general
expression for the fluctuations in the presence of a time
dependent voltage, valid for arbitrary relation between voltage
and temperature. Two limits are then treated: a pulse of voltage
and a DC voltage. A pulse of voltage causes phase sensitive
current fluctuations for which we derive microscopically an
expression periodic in $\int V(t)dt$ with the period  $h/e$.
Applied to current fluctuations in Josephson circuits caused by phase
slips, it gives an anomalous contribution to the noise with a
logarithmic singularity near the critical  current.  In  the  DC
case, we  get  quantum  to classical shot noise reduction factor
$1/3$, in agreement with recent results  of Beenakker and B\"uttiker.

\end{abstract}
\narrowtext

\noindent
{\it Introduction.}
Recently, the Landauer approach to electric
transport~\cite{Landauer} was extended to describe current
fluctuations~\cite{Lesovik,YK,Buttiker}. The central result,
formulated for a single channel conductor with transmission coefficient
$T$, is that at zero temperature the current noise magnitude is
given by $(e^2/h)T(1-T)eV$, where $V$ is the drop of voltage across
the system. Thus the quantum noise comes to be a factor
$1-T$ below the classical shot noise level $eI$. Beenakker and
B\"uttiker~\cite{BeenakkerButtiker} generalized this picture to a mesoscopic
conductor, where there are many conducting channels with a
distribution of transmission constants $T_n$. The noise is a sum
of contributions of separate channels: $\sum_n T_n(1-T_n)e^3V/h$.
Since in the limit of large dimensionless conductance $G$ the distribution of
$T_n$ is provided by the random matrix theory in the universal
form, $P(T)dT=Gdx$, where $T=1/\cosh^2x$, one obtains for the
noise ${1\over 3}Ge^3V/h$, i.e. the theory predicts universal
quantum to classical noise ratio $1/3$. Another approach developed
recently by Nagaev, who used kinetic equation\cite{Nagaev} and with
this technique also arrived at the factor $1/3$.

In order to understand better the relation of these results
with the conventional many-body methods, it would be of interest to do
a microscopic calculation. In this paper, we
study noise using non-equilibrium Keldysh technique and derive a
more general formula for current fluctuations caused by a time
dependent voltage. In the DC limit our results agree with those
found by other methods, and give a generalization to an AC
situation. We study current-current correlation
$\langle\!\langle I(t_1) I(t_2)\rangle\!\rangle$ in a mesoscopic
conductor under a slowly varying voltage and show that it is a
periodic function of $\Phi_{12}=c\int_{t_1}^{t_2}V(t)dt$ with
the period $\Phi_0=hc/e$, the single electron flux quantum. With
this, we treat current fluctuations in a metal caused by a short
pulse of voltage, or, equivalently, by a varying magnetic flux.
In this case the current fluctuations show phase sensitivity
(nonstationary Aharonov-Bohm effect) lacking in the DC
case~\cite{LesovikLevitov}. Recently, for a single channel
conductor geometry, the phase sensitive noise was expressed
through the transmission coefficient $T$ \cite{LeeLevitov}.
We
derive it below for a mesoscopic conductor within Keldysh
formalism, and show that the magnitude is given by the normal
metal conductance $G$ reduced by the same factor $1/3$.

We do the calculation for a cylindrical contact between two ideal
leads. As a simplest mesoscopic point of
view, we assume diffusive regime inside the contact and
completely ignore phase breaking effects, except for
the temperature. Also, we treat electrons as noninteracting fermions.
This somewhat restricts the range of applicability of our results,
since in real systems interactions are important. However, even for
interacting fermions, our calculation remains valid as long as
the Fermi liquid picture holds.

\smallskip
\noindent
{\it Main results.}
For the calculation we take two arbitrary sections $x_1$, $x_2$
of the contact, and find the current-current correlator
$S_{x_1,x_2}(t_1,t_2) = \langle\!\langle \hat I_{x_1}(t_1)\hat
I_{x_2}(t_2)\rangle\!\rangle$. It will be explicit in the
calculation that $S_{x_1,x_2}(t_1,t_2)$ does not depend on the
choice of the sections $x_1$, $x_2$, when the time scale
$\tau^\ast$ on which the voltage $V(t)$ varies is much longer
than the time of diffusion through the contact. Under the
assumption of large $\tau^\ast$, for a cylindrical
contact our result reads
  \begin{equation}\label{first}
S(t_1,t_2) =S_{eq}(t_1-t_2)\ \Bigl(\ 1-\frac{1}{6}
|1-e^{i\phi_{12}}|^2\ \Bigr)\ ,\qquad
\phi_{12}={2\pi c\over\Phi_0}\int\limits_{t_1}^{t_2}V(t)dt\ ,
  \end{equation}
where $S_{eq}(t_1-t_2)$ is the correlation in equilibrium:
  \begin{equation}\label{equilibrium}
S_{eq}(t)=\int e^2G\omega\coth{\omega\over2T}
e^{-i\omega t}{d\omega\over4\pi^2}= -{\rm Re}
{e^2GT^2\over2\sinh^2\pi T(t+i0)}\ .
  \end{equation}
The  correlator
exhibits Fermi anticorrelation in the time  domain,
since  it  is  explicitly negative at any $t_1\ne t_2$. Let us emphasize,
however, that  the  corresponding fluctuation of transmitted charge
is  positive,  due   to   a
compensation with a singularity in $S_{eq}(t_1,t_2)$ at $t_1=t_2$. Also,
since the second term of Eq.(\ref{first}) gives positive contribution
to  the  correlator, the inequality holds: $S(t_1,t_2)\ge
S_{eq}(t_1-t_2)$,  which means that the excess
noise is stricktly positive.

At constant  applied  voltage, $\phi_{12}={e\over\hbar}V(t_2-t_1)$,
the correlator $S(t_1,t_2)$ is a function only of $t_1-t_2$, and
thus it can be characterized by a spectral density:
 \begin{equation}\label{spectrum}
S_\omega={1\over6} (4 S^{eq}_{\omega}+S^{eq}_{\omega+eV}+S^{eq}_{\omega-eV})\ ,
\end{equation}
  where  $S^{eq}_\omega=e^2G\omega\coth{\omega\over2T}$  is  the
equilibrium  Nyquist noise  spectrum.  Then,  the noise at low frequency
$\omega\ll T,eV$ is given by
$ S_{0}={1\over 3}e^2G( 4T + \coth{eV\over2T} ) $.
In the limit $T=0$ this gives $S_0={1\over3}e^2GV$,
the celebrated quantum shot noise
result~\cite{BeenakkerButtiker,Nagaev}.

Beyond the DC situation, Eq.(\ref{first}) allows to study noise
in any AC setup, e.g., current fluctuations due to varying
magnetic flux or due to a pulse of voltage. Physically, in this
case the system can be realized as a normal metal ring in an AC
magnetic field, or as a shunt resistor of a superconducting
circuit in the regime of the nonstationary Josephson effect. We
consider a step-like time dependence of the flux, corresponding
to a pulse of voltage, and for $T=0$ derive an expression for the
fluctuations of transmitted charge:
  \begin{equation}\label{f4}
\langle\!\langle\delta Q^2 \rangle\!\rangle = \frac{1}{3}e^2G \left(
\frac{\Phi}{\Phi_0}+\frac{2}{\pi^2}\sin^2\pi{\Phi\over\Phi_0}
\ln\frac{t_0}{\tau^\ast}
\right),
  \end{equation}
where $\delta Q = \int_{-t_0}^{t_0}\hat I(t')dt' $ is the charge
transmitted over the interval $-t_0<t<t_0$, $\Phi$ is the height
of the flux step, and $\tau^\ast$ is the duration of the step,
assumed to be much shorter than $t_0$. A similar expression was
derived recently for a single channel
conductor~\cite{LeeLevitov}. In Eq.(\ref{f4}) the first term
corresponds to the $\omega=T=0$ noise (\ref{spectrum})
integrated over time, since the flux and the voltage are
related, $V(t)=-{1\over c}\dot\Phi(t)$. The second
$\Phi_0-$periodic term with an infrared logarithmic divergence
corresponds to the non-stationary AB
effect~\cite{LesovikLevitov}. We propose to observe it in a
Josephson circuit with a shunt, where it causes an anomalous
contribution to the noise in the shunt, diverging as $I$
approaches $I_c$.

\smallskip
\noindent
{\it General formalism.}
Let us turn to the calculation. In the current-current correlator
\mbox{ $\langle\!\langle\vec j(\vec r_1, t_1)
\vec j(\vec r_2, t_2) \rangle\!\rangle$ }
we take the times $t_1$ and $t_2$ on different branches
of the Keldysh contour, and write
$\langle\!\langle j_{1} j_{2} \rangle\!\rangle$ through the
functions $G^{+-}$ and $G^{-+}$ as
$\langle\!\langle j_1j_2\rangle\!\rangle =
{\rm tr}( j_1G^{+-}_{12}j_2G^{-+}_{21})$.
The trace ${\rm tr}$ means integration over inner momenta and energy,
summation over spin indices and averaging over configurations
of the random potential.
After expressing Fourier transform of the correlator
in terms of retarded and advanced Green's functions $G^R$ and $G^A$,
and the Keldysh function $F$, we get
  \begin{equation}\label{f6}
{\langle\!\langle jj \rangle\!\rangle }_{k,\omega} =
\frac{1}{4} {\rm tr}( j_{k,\omega} F j_{-k,-\omega} F )+
\frac{1}{2} {\rm tr}( j_{k,\omega} G^{R} j_{-k,-\omega} G^{A} ).
  \end{equation}
The functions $G^{R(A)}$ are familiar:
$G^{R(A)}_{\epsilon,p} =(\epsilon - \xi_p \pm i/2\tau)^{-1}$.
The function $F$ satisfies Dyson's equation~\cite{Keldysh}.
With the condition of short mean free path $l\ll L$,
it is reduced to
the diffusion equation for the quantity
${i\over \tau}\tilde s(r,t_1,t_2)$ defined according to
$F=G^R\tilde s-\tilde sG^A={i\over \tau}G^R\tilde sG^A$.
Following the usual scheme~\cite{Altshuler,LarkinKhmelnitskii},
we treat the vertex $\tilde s$ as a two-time diffuson
{}~\cite{LarkinKhmelnitskii,AltshulerAronov} that satisfies the
equation
   \begin{equation}\label{f17}
(\partial_{t_+}-D\tilde{\nabla}^2)\tilde s(\vec r,t_1,t_2)=0,\qquad
t_+=\frac{1}{2}(t_1+t_2)\ ,
  \end{equation}
together with the condition on the boundary with the leads :
   \begin{equation}\label{boundary}
\tilde s(\vec r, t_1,t_2)=\frac{i}{\tau}\tilde s_0(t_1-t_2)\ ,
  \end{equation}
where $\tilde s_0(t)=\int e^{-i\epsilon t} \tanh{\epsilon\over 2T}\
{d\epsilon\over2\pi}$,
i.e., the leads serve as reservoirs of equilibrium electrons.
The vector potential
$A(\vec r,t)$ enters Eq.(\ref{f17}) through
$\tilde \nabla = \nabla -
i\frac{e}{c}A(\vec r, t_1) + i\frac{e}{c}A(\vec r, t_2)$.
Initially, at $t_+=-\infty$,
$\tilde s(\vec r,t_1,t_2)=\tilde s_0(t_1-t_2)$ everywhere in the contact.
Note, that our definition of the
diffuson differs % by the factor $\frac{i}{\tau}\tilde s$
from that of
{}~\cite{AltshulerAronov} because $\tilde s$ is the vertex of the
function $F$ and thus it is a solution of the kinetic equation,
which in this case is reduced to Eq.(\ref{f17}) in a
conventional way.

Before we discuss averaging over disorder, let us evaluate the
simplest diagram shown in Fig.1(a). This contribution to the
current correlation is local, since it decays at distances $\gg l$.
Thus we write it through the conductivity $\sigma$ as
  \begin{equation}\label{source}
{\langle\!\langle\delta j^\alpha (r_1,t_1)
\delta j^\beta (r_2,t_2)\rangle\!\rangle} =
-{1\over2}e^2\sigma
{\rm Re}\left[\tilde s(r, t_1+i0,t_2) \tilde s( r, t_2-i0, t_1 ) \right]\
\delta_{\alpha\beta}\ \delta(r_1-r_2) ,
  \end{equation}
with the regularization at equal times that
follows from the second term of Eq.(\ref{f6}).
Now, we can  dress any current vertex of Fig.1(a)
by a diffusion ladder, which gives the other three diagrams of Fig.1.
Analytic elements are standard: the factor
$-iD\vec k e\pi\nu_0\tau$ corresponds to the current
vertex and the factor $(\pi\nu_0\tau)^{-1}/(-i\omega\tau+Dk^2\tau)$
corresponds to the diffuson. In the limit
$\tau^\ast\gg L^2/D$ one sets $\omega=0$ in the diffusons, and
then the sum of all contributions to the noise, the non-local
(Fig.1(b,c,d)) and the local (Fig.1(a)), is given by
  \begin{equation}\label{projector}
{\langle\!\langle j^\alpha (t_1) j^\beta (t_2)\rangle\!\rangle}_k =
\Bigl[\delta_{\alpha\alpha'}-{k_\alpha k_{\alpha'}\over k^2}\Bigr]
{\langle\!\langle\delta j^{\alpha'}(t_1)\delta
j^{\beta'}(t_2)\rangle\!\rangle}_k
\Bigl[\delta_{\beta\beta'}-{k_\beta k_{\beta'}\over k^2}\Bigr]
  \end{equation}
One  verifies that, due to the projector form of the expressions
in the brackets $[...]$, the correlator of currents through  two
arbitrary sections is independent of the choice of the sections.

The result (\ref{projector}) allows a simple and natural
interpretation in terms of a diffusion equation with random
current source,  ${\partial n\over\partial t}=D\nabla^2n-\nabla\delta j$.
Given Eq.(\ref{source}) for the fluctuations of $\delta j$,
one gets an expression for the fluctuation of
the total current $j=-D\nabla n+\delta j$ equal to the sum of
the graphs of Fig.1, which in the low frequency limit is reduced
to Eq.(\ref{projector}).

\smallskip
\noindent
{\it Cylindrical geometry.}
The rest of our discussion depends on the specific shape of the
contact. Let us consider a cylindrical contact of length $L$
between two ideal leads that serve as reservoirs of equilibrium
electrons.

Let us solve Eq.(\ref{f17}) assuming that
the  time  scale  $\tau^\ast$ on which the field varies is
longer than the diffusion time, $\tau^\ast \gg L^2/D $:
  \begin{equation}\label{f18}
\tilde s(x,t_1,t_2)  =  \tilde s_0(t_2-t_1)\left[
1-\frac{x}{L}+
\frac{x}{L}e^{i\phi(t_2)-i\phi(t_1)}
\right] \ e^{-i\phi(x,t_2)+i\phi(x,t_1)},
  \end{equation}
where $x$ is the coordinate along the cylinder axis, $0<x<L$,
$\phi(x,t)=\frac{2\pi}{\Phi_0}\int_{-\infty}^{x}
A(x',t)dx' $ and $\phi(t) = \phi(L,t)$. The solution obeys
boundary conditions (\ref{boundary}) at $x=0$ and $x=L$. With
$\tau^\ast \gg \frac{L^2}{D}$, one comes to Eq.(\ref{f18}) by
neglecting the time derivative of $\tilde s(\vec r,t_1,t_2)$ in
Eq.(\ref{f17}) with respect to the space derivative. Then the
vector potential in Eq.(\ref{f17}) can be eliminated by a gauge
transformation, where it should not be forgotten that the gauge
transformation changes the boundary conditions for the new
function $\tilde s'$: ${\left.\tilde s'\right|}_{x=L} =
\tilde s_0(t_2-t_1)e^{i\phi(t_2) - i\phi(t_1)}$.
Transformed function $\tilde s'$ satisfies the usual Laplace's
equation which is readily solved.

In the cylindrical geometry the quantity $k^{-2}$ in Eq.(\ref{projector})
should be interpreted as the Green's function
${\cal D}(\vec r_1,\vec r_2)$ of the Laplace's operator $\nabla^2$ .
According to Eq.(\ref{f18}), the source fluctuations (\ref{source}) are
homogeneous in every section of the cylinder, i.e., they depend
only on $x$. Thus the problem becomes effectively one dimensional,
and we can write :
  \begin{equation}\label{1d}
{\cal D}(x_1,x_2)_{\omega=0}=\cases{x_1(L-x_2)/L,\qquad x_1<x_2\cr
x_2(L-x_1)/L,\qquad x_2<x_1\cr}\ .
  \end{equation}
We substitute $\partial_{x_1}{\cal D}(x_1,x_2)\partial_{x_2}$
for $k^xk^x/k^2$ in Eq.(\ref{projector}), and readily
get the result (\ref{first}).

\smallskip
\noindent
{\it AC voltage noise.}
Now, we shall consider the fluctuations caused by a pulse of voltage of
duration $\tau^\ast\ll\hbar/T$. By setting $T=0$ in Eq.(\ref{first}) we get
  \begin{equation}\label{f20}
S(t_1,t_2) =-\frac{1}{3}\ e^2G\
{\rm Re}\ \frac{{|1-e^{i\phi_{12}}|}^2}{4\pi^2{(t_2-t_1+i0)}^2}\ .
  \end{equation}
Let us calculate the fluctuation of charge transmitted
during the interval $-t_0<t<t_0 $ :
$\langle\!\langle\delta Q^2 \rangle\!\rangle =
\int^{t_0}_{-t_0}\!\int^{t_0}_{-t_0}
S(t_1,t_2)dt_1dt_2$ .

The logarithmically diverging term of Eq.(\ref{f4})
is obtained from (\ref{f20})
by integrating over the times $t_1$ and $t_2$  before and after
the voltage pulse ($t_0/\tau^\ast\to\infty$)
  \begin{equation}\label{logterm}
{\langle\!\langle \delta Q^2 \rangle\!\rangle}_{log}  =
\frac {e^2G}{3\pi^2}\sin^2\frac{\pi\Phi}{\Phi_0}
\Biggl[ \int \limits_{-t_0}^{-\tau^\ast} dt_1 \int \limits_{\tau^\ast}^{t_0}
\frac{dt_2}{{(t_2-t_1)}^2} +
(t_1\leftrightarrow t_2)
\Biggr] =
\frac {e^2G}{3\pi^2}2\sin^2\frac{\pi\Phi}{\Phi_0}\ln\frac{t_0}{\tau^\ast}\ .
  \end{equation}
The contribution proportional to $\Phi/\Phi_0$
is extracted from almost coinciding times $t_1$ and $t_2$. For that,
we integrate (\ref{f20}) over $t={1\over2}(t_1+t_2)$
and $t'=t_2-t_1$. Assuming the time dependence of $\phi$ smooth
and monotonous, we write $\phi_{12} = \dot \phi(t)\,t'$ ,
do the integral over $t'$,
and with $\dot \phi(t) > 0$ come to the first term of
Eq.(\ref{f4}). At finite temperature,
Eqs.(\ref{logterm}),(\ref{f20}),(\ref{f4}) hold for
$t_0\le\hbar/T$, otherwise $t_0$ has to be replaced by $\hbar/T$.

Let us consider a Josephson junction shunted by a normal metal
resistor. If the current is fixed above the critical, $I>I_c$,
the voltage on the resistor oscillates with the Josephson
frequency $\omega=2e\bar V/\hbar$ :
  \begin{equation}\label{Josephson}
V(t)=R\frac{I^2-I^2_c}{I+I_c\cos \omega t},
  \end{equation}
with $R^{-1} = \frac{e^2}{h}G$ and $\bar V=R\sqrt{I^2-I^2_c}$.
At $I-I_c\ll I_c$, the signal
(\ref{Josephson}) corresponds to a periodic sequence of steps in
the flux, of the duration $\tau^\ast=RI_c$ and of the height
$\Phi={1\over2}\Phi_0$ each. Let us use Eq.(\ref{f4}) derived
for a single step to estimate the current noise in the shunt
caused by the signal (\ref{Josephson}). The step height $\Phi$
corresponds to the half-period of Eq.(\ref{f4}), and thus we get
a logarithmically diverging noise associated with each step. For
the low frequency noise spectrum this gives
  \begin{equation}\label{Jnoise}
S_{\omega=0}={1\over3}e^2G\bar V
\bigl\{1+{1\over2\pi^2}\ln{RI_c\over\bar V}\bigr\}\ ,
  \end{equation}
where the temperature is assumed to be small, $ T \ll e \bar V$.
The noise (\ref{Jnoise}) is anomalously large, since it exceeds
the shot noise level
${1\over3}e^2G\bar V$ by a logarithmic factor, diverging as $I\to I_c$.

Experimentally, the observation of the noise (\ref{Jnoise}) is
more feasible
in the temperature range $e\bar V<T<eRI_c$
{}~\cite{Koch}, where
it gives a correction to the thermal noise~\cite{Likharev}. For
such temperatures, $\ln{RI_c/\bar V}$ in Eq.(\ref{Jnoise})
should be replaced by $\ln{eRI_c/T}$.

\bigskip

We are grateful to G.B.Lesovik for illuminating discussions and useful remarks.
Research of L.L. is partly supported by Alfred Sloan fellowship.

\vfill\eject
\noindent {\bf Figure 1.}
Graphs for current fluctuations are shown: {\it (a)} the local;
{\it (b,c,d)} the non-local. Black vertices represent Keldysh function
${i\over\tau}\tilde s$ (see Eqs.(~\ref{f17}),(~\ref{boundary})), wavy
lines are diffusons, and each box is dressed by impurity lines to form
a Hikami vertex~\cite{Hikamibox}.

\end{document}